\RequirePackage{fix-cm}

\documentclass[twocolumn,epjc3]{svjour3}  

\RequirePackage{graphicx}
\smartqed  
                             
\usepackage{epsfig,dcolumn}
\usepackage{graphicx}
\usepackage{color}                 
\usepackage{amsmath}
\usepackage{amsfonts}
\usepackage{amssymb}
\usepackage{graphicx}
\usepackage{widetext}
\usepackage{type1cm}
\usepackage{eso-pic}

\newcommand{\be}{\begin{equation}}
\newcommand{\ee}{\end{equation}}

\newcommand{\im}{\mbox{Im}}
\newcommand{\re}{\mbox{Re}}

\usepackage{bm} 

\begin{document}

\title{The non-ordinary Regge behavior of the $K^*_0(800)$ or $\kappa$-meson 
versus the ordinary $K^*_0(1430)$.}

\author{J.R.~Pelaez
        \and
        A.Rodas	
}				
				
				\institute{Departamento de F\'isica Te\'orica II and UPARCOS, Universidad Complutense de Madrid, 28040 Madrid, Spain 
}


\maketitle

\begin{abstract}
The Regge trajectory of an elastic resonance can be calculated from dispersion theory,
instead of fitted phenomenologically, using 
only its pole parameters as input.
This also provides a correct treatment of resonance widths
in Regge trajectories, essential for very wide resonances. 
In this work we first calculate
 the  $K^*_0(1430)$ Regge trajectory, 
finding the ordinary almost real and linear behavior, typical of 
$q \bar q$ resonances. 
In contrast, for
the $K^*_0(800)$ meson, the resulting Regge trajectory is non-linear and 
has a much smaller slope than ordinary resonances, being remarkably
similar to that of the $f_0(500)$ or $\sigma$ meson.
The slope of these unusual Regge trajectories seems to scale with
the meson masses rather than with scales typical of quark degrees of freedom.
We also calculate the range of the interaction responsible for the
formation of these resonances. Our results 
 strongly support a non-ordinary, predominantly meson-meson-like, interpretation for the
lightest strange and non-strange resonances.
\end{abstract}
\maketitle

\section{Introduction}

There is  growing evidence for the existence of hadrons that do not follow
the ordinary quark-antiquark classification of mesons or the three quark classification of baryons. One of the most remarkable features of these ordinary resonances is the observation
that,
to a very good approximation, 
most hadrons can be {\it fitted}  into linear Regge 
trajectories with an almost universal slope of about 0.9 GeV$^{-2}$
in the spin versus mass squared plane. 
Regge trajectories are due to the analytic constraints of amplitudes in the 
complex angular momentum plane. Always subject to these constraints, 
different dynamics give rise to different
Regge trajectories relating the angular momentum and the parameters of poles associated to 
resonances. In particular, linear relations between the squared mass and the
angular momentum
are characteristic of relativistic confining interactions like a relativistic rotating rigid bar, flux tubes, string dynamics, etc. or those between quarks in QCD. The slope
of such linear trajectories is related to the ``string tension'' or energy density of the tube connecting the various quarks in the hadron. 
However, different dynamics do not necessarily lead to linear Regge trajectories. 

Let us remark that, due to the analytic properties of amplitudes in the complex plane, 
in certain cases Regge trajectories can be {\it calculated} from 
the properties of just one resonance \cite{Londergan:2013dza,Carrasco:2015fva}, {\it instead of fitted} to several resonances
assuming that a straight line should describe them. This 
approach relies on dispersion relations and unitarity constraints for Regge trajectories and residues
\cite{Reggeintro,Collins-PLB,Chu:1969ga}
and
is more fundamental and predictive than a pure straight line fit.
Actually, it does not assume a priori a particular functional form for the
trajectories.  In addition, it includes a consistent treatment of resonance widths, which are usually neglected in the usual phenomenological fits of Regge trajectories.
This allows for a clear identification of wide resonances, instead of using, incorrectly, the width as a source of uncertainty in the fits.

The method has been described and applied 
recently in \cite{Londergan:2013dza,Carrasco:2015fva}.
On the one hand, four ordinary linear Regge trajectories were found
from the $\rho(770)$, $f_2(1270)$, $f'_2(1525)$ and $K^*(892)$ resonance poles. 
The slopes obtained are fairly close to 0.9 GeV$^{-2}$, 
expected to be universal for all ordinary trajectories.
This is just a confirmation of their well-established ordinary nature,
although the observed partners in their Regge trajectories can be understood as predictions
of the approach, since the input only comes from the above four particles.
On the other hand, for the controversial $f_0(500)$ or $\sigma$ meson, 
whose position has recently been determined accurately in \cite{Caprini:2005zr} using rigorous 
and model-independent dispersive formalisms,
a non-linear Regge behavior with a much smaller slope and a much larger imaginary part was found \cite{Londergan:2013dza}. 
Its Regge trajectory was strikingly similar to that of a Yukawa potential, at least
below 2 GeV$^2$. Moreover, by imposing a linear Regge trajectory on the $\sigma$ pole position
the dispersion relations yielded an amplitude that was at odds with the scattering data, even qualitatively. This justified the omission of the $f_0(500)$
from $(J,M^2)$ ``quarkonia'' linear fits in \cite{Anisovich:2000kxa},
and it provided strong support for the generally accepted 
non-ordinary nature of this meson, which may contain a large, or even dominant,
 meson-meson component (see \cite{Pelaez:2015qba} for a recent review).

In this work, after briefly reviewing the method in the next section,
we extend this research further into the scalar strange sector in Sect. \ref{sec:numerical results}.
As a further check of the reliability of the approach in the strange sector, we study first
the   $K^*_0(1430)$ in Sect.~\ref{subsec:K0(1430)}.
The elastic formalism is a good approximation because, following the Review of Particle Properties (RPP) \cite{RPP}, 
the $K^*_0(1430)$ branching ratio to $K\pi$ is $(93\pm 10\%)$.
Our calculation gives rise to 
an almost real and linear Regge trajectory, 
with a slope very consistent with the universal value.

In contrast, in Sect.~\ref{subsec:K0(800)} 
we show that the controversial $\kappa$ or $K^*(800)$ meson
results in a non-ordinary trajectory, whose imaginary part is
larger than the real part, which is not linear and whose slope 
is much smaller than the universal slope of ordinary trajectories. 
This is a new piece of evidence supporting the non-quark-antiquark nature of this state (tetraquark, meson-meson ``molecule'', different admixtures of these, etc)
which has been suggested from many other approaches 
\cite{Jaffe:1976ig,vanBeveren:1986ea,Oller:1997ng,Black:1998zc,Oller:1998zr,Close:2002zu,Pelaez:2003dy,Pelaez:2004xp}.
The existence of this state has been the subject of a long debate and it is still listed under ``Needs confirmation'' in the RPP. However, by means of a rigorous 
dispersive analysis in \cite{DescotesGenon:2006uk}, based on Roy-Steiner partial wave dispersion relations,
it has been confirmed that a pole associated to that state exists. This pole, below 800 MeV
 is found by many other approaches based on chiral symmetry and/or dispersion relations 
\cite{Oller:1997ng,Black:1998zc,Oller:1998zr,Pelaez:2004xp,Wolkanowski:2015jtc},
in analytic extractions of poles without model-dependent assumptions \cite{Cherry:2000ut,Pelaez:2016klv} or very recently on the lattice \cite{Dudek:2014qha} (although given the high quark
masses used in the calculation it appears as a virtual state, as suggested in \cite{Nebreda:2010wv}
from dispersion theory and effective chiral Lagrangians).

Moreover, in Sect.~\ref{subsec:K0(800)} we will discuss the striking similarities of the $\kappa$ trajectory calculated
at low energies with 
the trajectory of the $\sigma$ meson as well as with Yukawa potentials.
In particular, we show that the range of a Yukawa potential 
that would mimic the trajectories of this resonances seems to scale with the
reduced mass of the system, suggesting an important role for meson-meson dynamics
in the formation of these resonances. 
The range of this Yukawa potential is a well-defined and intuitive
measure of the scale involved in the $\sigma$ and $\kappa$ formation, in contrast to the 
conventional mean-squared radius, which is ill-defined for resonances since they are non-normalizable states. The spatial scale of a resonance is of interest to discuss its nature as a composite or compact object, and the scales we find are somewhat smaller but comparable to typical meson-meson scattering lengths.
In Section~\ref{conclusions} we will present our conclusions.

\section{Dispersive calculation of Regge trajectories}

Following \cite{Londergan:2013dza,Carrasco:2015fva}, let us briefly recall the notation and the 
derivation of the dispersion relations that determine the
Regge trajectory and residue of an elastic resonance just from its pole parameters. 
The partial 
wave expansion of the kaon-pion scattering amplitude $T(s,t)$ is
\be
 T(s,t)=32 \pi \sum_l (2l+1) t_l(s) P_l(z_s(t)),
\label{fullamp}
\ee
where $z_s(t)$ is the cosine of the s-channel scattering angle. 
In the elastic region the partial waves can be parameterized as
\begin{align}
&t_l(s) =  e^{i\delta_l(s)}\sin{\delta_l(s)}/\rho(s), \quad \rho(s) = 2q(s)/\sqrt{s},
\\ &q(s)=\sqrt{(s-(m_K+m_\pi)^2)(s-(m_K-m_\pi)^2)/4s},
\end{align}
where $l$ is the angular momentum, $\delta_l(s)$ is the phase shift
and $q(s)$ is the center-of-mass momentum. Thus that the partial wave has a
branch cut from threshold to infinity.
Near the pole of a resonance with spin $l$ the partial wave reads
\be
t_l(s)  = \frac{\,\beta(s)}{l-\alpha(s)\,} + f(l,s),
\label{reggepw}
\ee
where $f(l,s)$ is an analytic function around $l=\alpha(s)$. 
The complex function $\alpha(s)$ is called the Regge trajectory 
of the resonance and $\beta(s)$ its residue. Both functions satisfy the Schwartz reflection
symmetry also satisfied by the partial wave, {\it i.e.}, $\alpha(s^*)=\alpha^*(s)$ and $\beta(s^*)=\beta^*(s)$. If we now  consider 
a region where the pole dominates the partial wave behavior, then the unitarity condition $\mbox{Im}t_l(s)=\rho(s)|t_l(s)|^2$ implies that
\be
\mbox{Im}\,\alpha(s)   = \rho(s) \beta(s).   
\label{uni} 
\ee
Moreover, the elastic unitarity condition determines the analytic continuation
of $t_l(s)$ through the cut to the second Riemann sheet, where resonance poles may occur. Similarly, Eq. \eqref{uni} determines the analytic continuation 
of $\alpha(s)$ \cite{Chu:1969ga}. 

Let us now recall that near threshold partial waves behave as $t_l(s) \propto q^{2l}$, therefore if the resonance pole dominates 
the partial wave, then  $\beta(s) \propto q^{2\alpha(s)}$ in that region. Moreover the Regge contribution to the amplitude is proportional to $(2\alpha + 1) P_\alpha(z_s)$, hence, in order to cancel 
the spurious pole of the  Legendre function 
$P_\alpha(z_s)\propto\Gamma(\alpha + 1/2)$. 
The residue must vanish in that region of energy whenever $\alpha(s) + 3/2$ is a negative integer, {\it i.e.}, it is convenient to write
\be
\beta(s) =  \gamma(s) \hat s^{\alpha(s)} /\Gamma(\alpha(s) + 3/2),  
\label{betaexpre} 
\ee
where $\hat s =4 q^2/s_0$ 
and in order to have the right dimensions, we have introduced a scale $s_0$, which we 
conveniently set to $s_0=1\,$ GeV$^2$ without losing generality. 
The so-called reduced residue, $\gamma(s)$, is a real 
analytic function. Hence, on the real axis above threshold, since $\beta(s)$ is real, the phase of $\gamma$ is
\be
\mbox{arg}\,\gamma(s) = - \mbox{Im}\alpha(s) \log(\hat s) + \arg \Gamma(\alpha(s) + 3/2). 
\ee
Consequently, we can write for $\gamma(s)$ a dispersion relation using an Omn\'es function:
\be
\gamma(s) = P(s) \exp\left(c_0 + c' s + \frac{s}{\pi} \int_{(m_K+m_\pi)^2}^\infty \!\!\!\!ds' \frac{\mbox{arg}\,\gamma(s')}{s' (s' - s)} \right), \label{g}
\ee
where $P(s)$ is an entire function. 
The large-$s$ behavior is not determined from first principles, but linear Regge trajectories 
are expected for ordinary mesons and thus we  
allow $\alpha$ to behave as a first order polynomial at large-$s$. 
Thus we only need to use 
two subtractions to obtain a dispersion relation~\cite{Reggeintro,Collins-PLB}: 
\be
\alpha(s) = \alpha_0 + \alpha' s + \frac{s}{\pi} \int_{(m_K+m_\pi)^2}^\infty ds' \frac{ \mbox{Im}\alpha(s')}{s' (s' -s)}. 
\label{alphadisp}
\ee
Let us remark that in \cite{Carrasco:2015fva} it was shown that considering three subtractions lead
to almost indistinguishable results.
Therefore linear trajectories are not imposed a priori and, actually, a non-linear behavior was
found for the $f_0(500)$ resonance \cite{Londergan:2013dza}.

From Eq. \eqref{uni} it then follows that $c' = \alpha' ( \log(\alpha'  s_0) - 1)$ and that $P(s)$ can only be a constant. Therefore, we arrive at the following equations
\cite{Chu:1969ga,Londergan:2013dza,Carrasco:2015fva}
describing the Regge trajectory of a resonance pole when it dominates
a partial wave as in Eq. \eqref{reggepw}:

\begin{widetext}
\begin{align}
\mbox{Re} \,\alpha(s) & =   \alpha_0 + \alpha' s +  \frac{s}{\pi} PV \int_{(m_K+m_\pi)^2}^\infty ds' \frac{ \mbox{Im}\alpha(s')}{s' (s' -s)}, \label{iteration1}\\
\mbox{Im}\,\alpha(s)&=  \frac{ \rho(s)  b_0 \hat s^{\alpha_0 + \alpha' s} }{|\Gamma(\alpha(s) + \frac{3}{2})|}
 \exp\Bigg( - \alpha' s[1-\log(\alpha' s_0)] 
+  \!\frac{s}{\pi} PV\!\!\!\int_{(m_K+m_\pi)^2}^\infty\!\!\!\!\!\!\!ds' \frac{ \mbox{Im}\alpha(s') \log\frac{\hat s}{\hat s'} + \mbox{arg }\Gamma\left(\alpha(s')+\frac{3}{2}\right)}{s' (s' - s)} \Bigg), 
\label{iteration2}\\
 \beta(s) &=    \frac{ b_0\hat s^{\alpha_0 + \alpha' s}}{\Gamma(\alpha(s) + \frac{3}{2})} 
 \exp\Bigg( -\alpha' s[1-\log(\alpha' s_0)] 
+  \frac{s}{\pi} \int_{(m_K+m_\pi)^2}^\infty \!\!\!\!\!\!\!ds' \frac{  \mbox{Im}\alpha(s') \log\frac{\hat s}{\hat s'}  + \mbox{arg }\Gamma\left(\alpha(s')+\frac{3}{2}\right)}{s' (s' - s)} \Bigg),
 \label{betafromalpha}
 \end{align}
where $PV$ denotes the principal value.  For real $s$, the last two equations reduce to Eq.\eqref{uni}.
\end{widetext}

The dispersive approach to calculating Regge trajectories
consists on solving those three equations  
numerically with the free parameters fixed by demanding 
 that the pole on the second sheet of the amplitude in Eq.~(\ref{reggepw}) 
reproduces the position and residue of the pole associated to the resonance under study.
As already commented in the introduction this procedure yields almost real and linear Regge trajectories with a universal slope of $\sim $0.9 GeV$^{-2}$ for the $\rho(770)$ \cite{Londergan:2013dza}
 $f_2(1270)$, $f'_2(1525)$ and $K^*(892)$ resonances \cite{Carrasco:2015fva}. In contrast it leads to a very unusual non-linear trajectory for the $f_0(500)$ or $\sigma$ meson.

For scalars like the $f_0(500)$, studied in \cite{Londergan:2013dza}, or the $K^*_0(800)$ and the $K^*_0(1430)$, which will be studied here,
 the method is slightly modified \cite{Londergan:2013dza} to factor out explicitly
in the residue
the Adler-zero of the partial wave required by chiral symmetry, 
namely, $\beta(s)\propto(s-s_A)$.
For our purposes here it is enough to place it at its leading order position within 
Chiral Perturbation Theory, which for kaon-pion scattering is at $s_A=0.236\;$GeV$^2$. 
Then, since we do not want to spoil the large $s$-behavior, we need to replace 
$\Gamma(\alpha+3/2)$ by $\Gamma(\alpha+5/2)$. A spurious pole appears now
at $\alpha=-3/2$, but this is far away from the resonance region and 
hence becomes irrelevant from the calculation. 
In summary, for the $K^*_0(800)$ 
and $K^*_0(1430)$ the right hand side of Eq. \eqref{betafromalpha} 
should contain an $(s-s_A)$ factor 
and all instances of $3/2$ in the $\Gamma$ functions should be replaced by  $5/2$. Then $b_0$ has GeV$^{-2}$ dimensions.

Before discussing our numerical results 
let us recall the relation between the coupling $g$ of the resonance to 
its dominant decay channel and the residue of the pole $\vert Z\vert$:
\begin{equation}
\vert g\vert^2=\frac{16\pi(2l+1)|Z|}{\vert 2q(s_p)\vert^{2l}}.
\label{eq:coupling}
\end{equation}

Note that by calculating $\alpha(s)$ from the pole of one elastic resonance,
we are not only obtaining the real part of the
trajectory, which predicts the mass of the next partner in the trajectory, but also the imaginary part, 
which can be naively converted into a prediction of the width. 
In particular, for Breit-Wigner resonances that have just a single dominant decay mode, 
their width can be related to their Regge trajectory as $\Gamma={\text Im}\alpha/(M \re \alpha')$. 
There are several caveats here: first, that we have assumed a Breit-Wigner form.
Second, that the next partner of the linear trajectory found for the $K_0^*(1430)$
is heavy enough to lie beyond the strict applicability 
limits of our approach (the elastic or almost elastic region).
Thus, being obtained from the extrapolation of our results to high energies, this $\Gamma$ can only be 
considered as an estimate.
Third, that, being in the inelastic region, the partner does not have to 
decay predominantly into a single mode,
so that the $\Gamma$ above should only be interpreted as the partial width to $K\pi$. With these caveats in mind
we will see that the partial-width estimate is fairly reasonable.

In practice, it is the full
 elastic amplitude, including the background, the one that satisfies elastic unitarity.
Therefore, 
our approximation that the pole contribution alone satisfies elastic unitarity
is only valid in the region where the pole dominates the partial wave. 
However, dispersion relations are integrated from threshold to infinity. 
There are two possibilities now: to restrict the integrals to the region where the resonance pole dominates,
or to use the one-pole approximation in the whole energy region. In the results we describe in the next section
we have opted for the second
one but we have checked that the results change little if we use the first option. 
In particular, about 90\% of the integral comes from $s'$ within roughly one width of the resonance
in the $s$ region of interest (again within roughly one width of the resonance).
We will then compare our results in the surroundings of each resonance, where they are to be trusted, 
and cover with a mesh the areas where our approximation is not expected to hold.

\section{Numerical results}
\label{sec:numerical results}
 
Strictly speaking, the method described in the previous section
is suitable for resonances appearing in
the elastic scattering of two mesons. 
In the strange sector this is fulfilled by the vector $K^*(892)$, already studied in 
\cite{Carrasco:2015fva}, and the $K^*_0(800)$ to be studied below in Sect.~\ref{subsec:K0(800)},
since in practice both have a 100\% branching ratio to $K\pi$.
However, it was shown  in \cite{Carrasco:2015fva}
that the method is also able to reproduce the ordinary 
behavior of the $f_2(1270)$, $f'_2(1525)$ resonances, which are almost elastic, each with a dominant
decay  whose branching ratio is larger than 84\%.
For this reason, we are confident to extend the approach here to the 
$K^*_0(1430)$, whose branching ratio to $K\pi$ is $(93\pm 10) \%$ according to the RPP \cite{RPP}.
Thus, we will consider this small inelasticity 
as a source of systematic error and include an additional
 7\% uncertainty in the $K^*_0(1430)$ residue.

For each resonance
we obtain the best values for $\alpha_0, \alpha'$ and $b_0$ 
by fitting  the pole in Eq. \eqref{reggepw} to the parameters of the observed associated pole, 
where $\alpha(s)$ and $\beta(s)$ are 
numerical solutions of 
Eqs. \eqref{iteration1} and  \eqref{betafromalpha}.
Hence, the inputs to calculate each Regge trajectory 
are just the pole position  $s_p$ and residue $|g^2|$ of a single resonance.

In practice, at each step in the fitting procedure a set 
of $\alpha_0, \alpha'$ and $b_0$ parameters is chosen 
and the system of Eqs.\eqref{iteration1} 
and \eqref{iteration2} is solved iteratively. 
The resulting Regge amplitude for each $\alpha_0, \alpha'$ 
and $b_0$ is then continued to the 
complex plane to find a pole.
From this pole we define a $\chi^2$ function 
by calculating the differences between the mass, 
width and coupling observed values of the pole under study and
the pole obtained from the above equation, divided by the uncertainties. 
The best values for $\alpha_0, \alpha'$ and $b_0$ are 
obtained by minimizing this $\chi^2$ function.

\subsection{$K^*_0(1430)$ resonance}
\label{subsec:K0(1430)}
 
\begin{figure*}
\begin{center}
\includegraphics[scale=0.32]{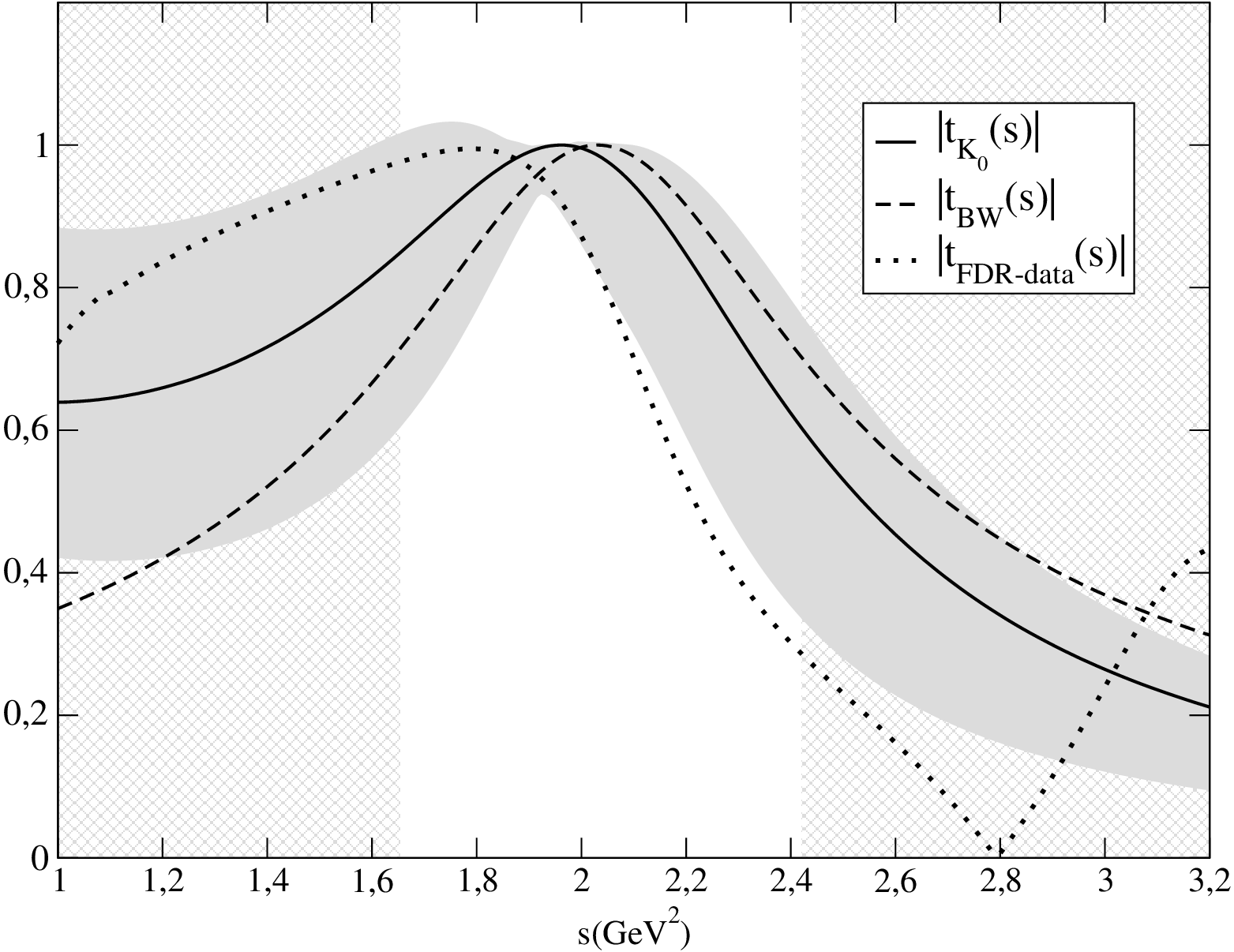}
\hspace*{.1cm}
\includegraphics[scale=0.32]{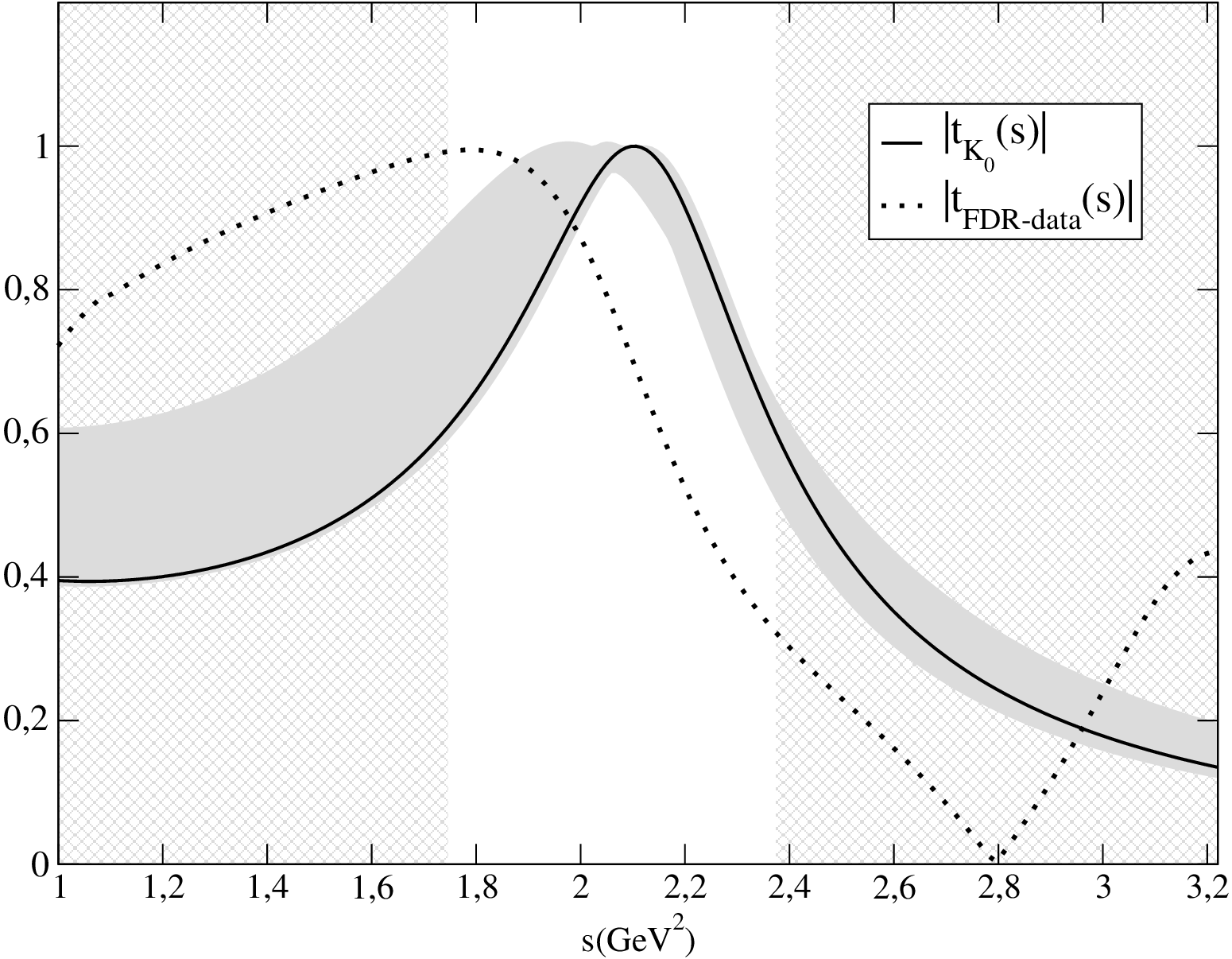}
\end{center}
 \caption{\rm \label{Fig:ampliK0} 
The dotted line, labeled ``FDR-data'', corresponds to the modulus of the $K\pi$ scattering partial wave
in the scalar-isospin 1/2 channel, obtained from a fit to data constrained with Forward Dispersion Relations \cite{Pelaez:2016tgi}. On the left panel, the dashed line is the Breit-Wigner shape obtained using the
resonance parameters in the RPP. Note that the Breit-Wigner peak is displaced from the peak in the data.
The continuous line is the result of our method, with a Regge pole consistent with that of the Breit-Wigner parameterization, 
but satisfying the dispersive constraints on the Regge trajectory Eqs.(\ref{iteration1}) to (\ref{betafromalpha}).
On the right panel we show similar results, but avoiding a Breit-Wigner or other particular model.
The resulting Regge pole shape is also somewhat displaced with respect to data but slighthly narrower than 
when assuming a Breit-Wigner formalism.
In both cases the gray bands cover the uncertainties due to the errors in the input pole parameters.
The regions covered with a mesh correspond to $s<(M-\Gamma/2)^2$ and $s>(M+\Gamma/2)^2$.}
\end{figure*}

According to the RPP, the $K^*_0(1430)$ is a well-established resonance, whose parameters are obtained both from 
$K\pi$ scattering and decay from production processes. Note, however, that our formalism is based on 
scattering amplitudes (particularly due to the use of the unitarity condition in Eq. \eqref{uni}, and
when looking at scattering data the nearest peak to 1430 MeV actually occurs at a somewhat lower energy.
In addition, even taking into account that it is only approximately elastic,
the amplitude does not follow a typical isolated Breit-Wigner shape. 
These two features can be seen in both panels of Fig.\ref{Fig:ampliK0}, where we have represented as a dotted line
the modulus of the amplitude obtained in 
a recent reanalysis \cite{Pelaez:2016tgi} of scattering data \cite{Kpiscattering} constrained to satisfy Forward Dispersion Relations (FDR).
The reason for such a behavior can be attributed to the presence of backgrounds, possibly from other resonances. In particular they may come
from the still controversial $K_0^*(800)$, whose width is of the order of 600 MeV, and maybe also
from  another still disputed $K_0^*(1950)$ resonance, with a with of the order of 200 MeV.
Thus, in this case, although the presence of the resonance is undisputed, 
there is some spread in its parameters, particularly on the width.

Thus, we are going to deal with the pole of the $K^*_0(1430)$ resonance following two different approaches.
Within the first, more conservative approach, we will use a very simple description using a Breit-Wigner (BW) functional
form. 
The parameters of this ``BW-pole'' are obtained from  the RPP 
and read
\begin{eqnarray}
 \sqrt{s_{K^*_0}}&=&M-i\Gamma/2=(1431\pm 50)-i(135\pm40)\text{ MeV},\nonumber\\ 
|g_{K^*_0}|^2&=&22.0\pm6.2\text{ GeV}^{2}.
\label{K1430BWparam}
\end{eqnarray}
The resulting BW line shape is shown on the left panel of Fig.\ref{Fig:ampliK0} as a dashed line.
Note that, despite dominating the amplitude in that region,  this BW form does not describe the data accurately, which implies the existence of a background.
Fortunately, for our approach only the pole parameters are needed.

Within the second approach we will use a recent pole determination \cite{Pelaez:2016klv} that does not assume a 
particular functional form or model for the pole, but uses a sequence of Pad\'e approximants
with powerful convergence properties in the complex plane. 
This sequence is calculated 
from the values of the amplitude
and its derivatives at an energy point near the resonance. The values of
the amplitude are taken from the recent analysis  of scattering data \cite{Pelaez:2016tgi}
constrained with forward dispersion relations.
This approach is meant to minimize the model dependence. In this case the ``PFDR-pole'' parameters are:
\begin{eqnarray}
\sqrt{s_{K_0^*}}&=&(1431\pm6)-i(110\pm19)\;{\rm MeV},\nonumber\\
|g_{K_0^*}|^2&=&14.6\pm5.6\;{\rm GeV^2}. 
\label{K1430Padeparam}
\end{eqnarray}

Still, one might be concerned about the description of data 
and try to get a  more accurate parameterization in terms of more Regge
poles. Actually, the parameterizations in \cite{Pelaez:2016tgi} do have several poles
 \cite{Pelaez:2016klv} and describe the data very accurately.
However, if one tries to implement a dispersive formalism with more Regge poles, 
each one has three more functions to determine 
(Re$\,\alpha$, Im$\,\alpha$ and $\beta$), but still just one elastic unitarity condition 
for the whole partial wave.
Thus, one does not obtain a closed system of integral equations.
It is only because we assume that elastic unitarity is good for each pole separately
that we can derive the powerful system of two integral relations provided in Sect. 2,
relating the real and imaginary parts of each pole trajectory.

Therefore the input for our equations are just the pole parameters of each resonance, 
and these have to be extracted by isolating each pole contribution, as we have just done in the previous paragraphs. 
The pole itself does not have to describe the data perfectly, 
since there is a background to complete the description.
For the trajectory of a given resonance 
only its own pole is relevant, everything else is background, no matter whether it comes from another resonance. In particular, 
the $K_0^*(800)$ and the $K_0^*(1430)$ are fairly well separated. Thus,
when extracting the parameters
of one of them, the contributions of the other one should be considered background.

Hence, we assume that $\beta$ and $\alpha$ are related by elastic unitarity.
For this approximation to hold, elastic unitarity should be a good approximation for $t_l$ (which is indeed the case as shown in \cite{Pelaez:2016tgi}) 
and the pole {\it should dominate} 
the partial wave in a certain region.
The method is then valid in that same region. 
This is why we provided the curves in Fig. \ref{Fig:ampliK0}, 
just to show that the pole contribution (extracted in \cite{Pelaez:2016klv}
 from a parameterization that describes the data accurately) 
dominates the amplitude in that same region.  
One might be worried that the peak is somewhat displaced, but 
we will also
make the calculation with a Breit-Wigner functional form, which by construction satisfies unitarity, and we will check that the results 
are compatible with those obtained from the pole extracted in \cite{Pelaez:2016klv}.

We then apply the method explained in the  previous section to these two determinations of the pole.
For the ``BW-pole'' approach, its values Eq. \eqref{K1430BWparam} are well fitted,
resulting in $\sqrt{s_{K^*_0}}=(1431\pm51)-i(139\pm65)\text{ MeV}$ and a coupling of $|g_{K^*_0}|^2=21.6\pm9.1\text{ GeV}^{2}$. The larger errors obtained for the width and the coupling are caused mostly by the systematic uncertainties included for the branching ratio, since it has a 7\% inelasticity. 
Then, on the left panel of Fig.~\ref{Fig:ampliK0} we show 
as a continuous line the Regge-pole amplitude resulting from our method.
We see that even though we have just fitted the pole, 
which is the only relevant feature for the Regge trajectory, 
this amplitude is rather similar to the Breit-Wigner form.
The gray bands cover the uncertainties in the Regge-pole amplitude arising from the errors of the input.

We follow the same steps for the ``PFDR-pole''. Its values in Eq. \eqref{K1430Padeparam} are well fitted,
resulting in $\sqrt{s_{K^*_0}}=(1431\pm6)-i(110\pm22)\text{ MeV}$ and a coupling of $|g_{K^*_0}|^2=15.0^{+5.3}_{-1.96}\text{ GeV}^{2}$. Once again the larger errors 
obtained for the width and the coupling are caused mostly 
by the estimation of systematic uncertainties due to the 7\% inelasticity.
This time we show on the right panel of Fig.\ref{Fig:ampliK0} the resulting
Regge-pole amplitude, whose peak is somewhat narrower than that of the Breit-Wigner shape in the left panel.
The gray bands cover the uncertainties in the Regge-pole amplitude arising from the errors of the input.

In the process of fitting  to the observed values the pole in Eq. \eqref{reggepw}, with the constraints in
 Eqs. \eqref{iteration1} to \eqref{betafromalpha}, 
we obtain the $b_0$, $\alpha_0$ and $\alpha'$ 
parameters.  
For the BW-pole they  are:
\begin{align}
&\alpha_0=-1.10^{+0.04}_{-0.21}\,;\hspace{0.3 cm}
\alpha'=0.78^{+0.07}_{-0.13} \text{ GeV}^{-2};\\ \nonumber
&b_0=4.08^{+1.08}_{-3.19}\text{ GeV}^{-2}\,,\label{eq:paramk0}
\end{align}
whereas for the PFDR-pole trajectory we find
 \begin{align}
&\alpha_0=-1.28^{+0.01}_{-0.17}\,;\hspace{0.3 cm}
\alpha'=0.81^{+0.01}_{-0.04} \text{ GeV}^{-2};\\ \nonumber
&b_0=2.5^{+1.1}_{-0.4}\text{ GeV}^{-2}\,.\label{eq:paramk0PFDR}
\end{align}

\begin{figure*}
\begin{center}
\includegraphics[scale=0.4]{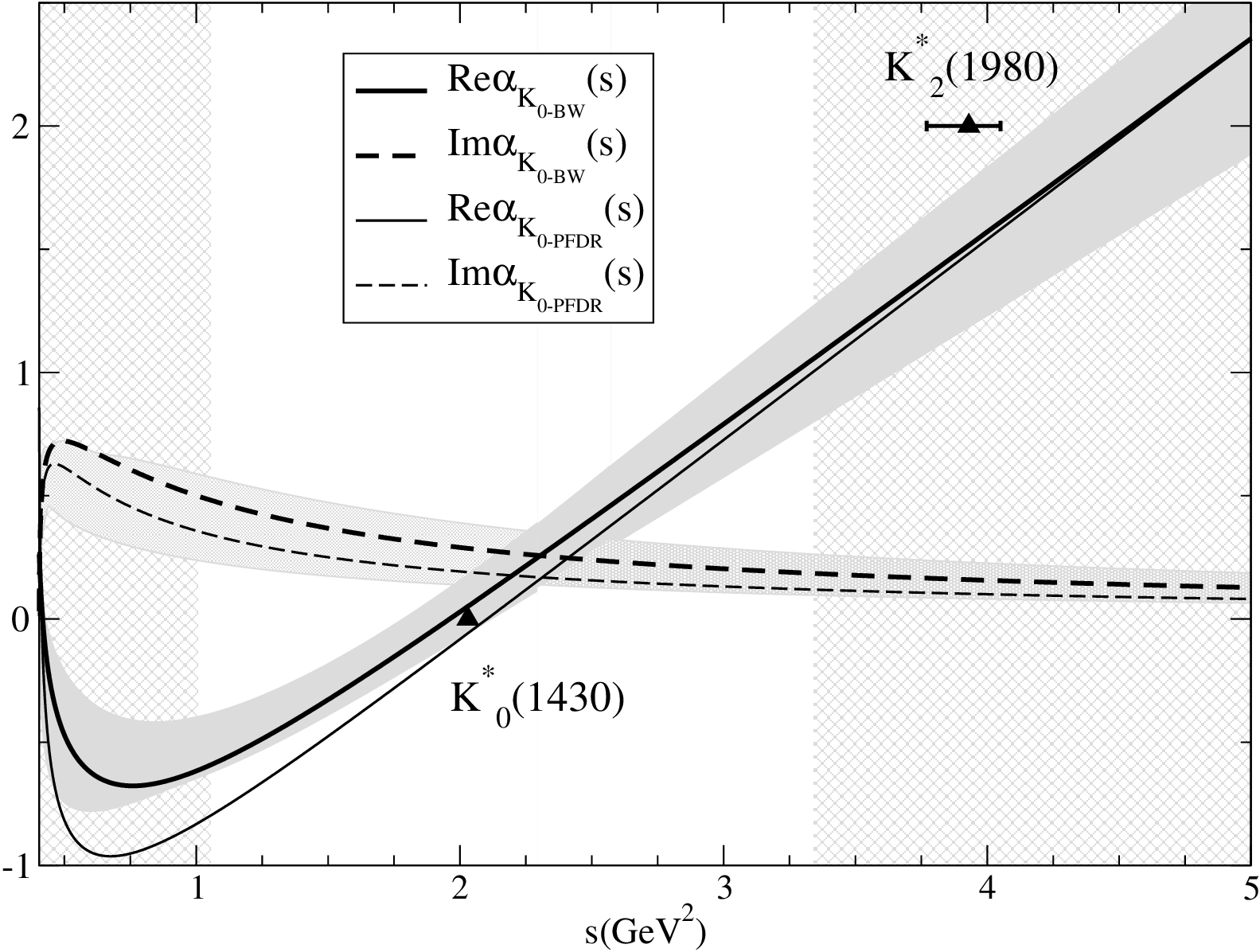}
\end{center}
 \caption{\rm \label{Fig:alphak0} 
Real (solid) and imaginary (dashed) parts of the $K^*_0(1430)$ Regge trajectory.
 The gray bands cover the uncertainties due to the errors in the input pole parameters for the Breit-Wigner pole, which central values are the thick lines, the values obtained for the PFDR pole are the thin lines.  The area 
covered with a mesh is the mass region starting
three half-widths above and below the resonance mass, where our 
approximation that $K\pi$ is elastic and dominated 
by the $K^*_0(1430)$ does not longer hold and where our approach should be considered only as a mere extrapolation. We show the $K_0^*(1980)$ resonance listed in the RPP 
that seems a good candidates for a $K_0^*(1430)$ partner in this trajectory.}
\end{figure*}

With these parameters the trajectory $\alpha(s)$ is fully determined as a solution of the integral equations. Thus, in Fig. \ref{Fig:alphak0} 
we show the resulting trajectories for the BW-pole (thick lines) and the PFDR-pole (thin lines).
The real part of the trajectories is shown as a continuous line and the imaginary part as a dashed line.
Both pole determinations yield very similar trajectories. 
In the figure we have covered with a light mesh the regions that lie beyond three half-widths of the
$K^*_0(1430)$ mass, where we do not expect our method to give accurate results and the curves should be considered a qualitative extrapolation. 
Within the applicability region, 
we find that the real part of the trajectory is almost linear and bigger in modulus than the imaginary part above the resonance mass. In other words, it comes out as expected for Regge trajectories of ordinary mesons.

Nevertheless, in Fig.~\ref{Fig:alphak0} we show the resulting Regge trajectory up to $s=5$ GeV$^2$. The reason for such an extrapolation is to show the position of the $K^*_2(1980)$ mass at  $M_{K^*_2(1980)}=(1973\pm8\pm25)\,$MeV \cite{RPP}, 
which could be the next state in the Regge trajectory. 
It should be noted that this resonance is listed in the RPP, but omitted from the summary tables, because
it ``Needs confirmation'' \cite{RPP}.
No other $J^P=2^+$ candidate is found nearby in the RPP, particularly not with a slightly higher mass. 
This resonance is fairly close to our extrapolated trajectory.
This can be considered as  further support for its existence.
However, it should be noted that it is somewhat lighter than expected from our results,
although  
one should take into account that the mass listed in the RPP is not the pole mass that we use in our calculations.

It is worth remarking that the $K^*_2(1980)$ is not only slightly off our trajectory,
but also off from typical linear trajectories. In particular, forcing 
 the $K_0^*(1430)$ and the $K_2^*(1980)$ to lie on the same straight trajectory $J=\alpha_0+\alpha' M^2$, yields a slope $\alpha'\simeq 1.07\,$GeV$^{-2}$, which is somewhat larger than the usual value of $0.9\,$GeV$^{-2}$. 
This small tension with the universal slope could be due 
to the fact that, although  the $K_0^*(1430)$ is generally 
accepted as an ordinary quark-antiquark meson, 
it might also have some small mixing with other
meson configurations \cite{Fariborz:2005gm,Giacosa:2006tf}. 
This would be rather natural since such non-ordinary mesons candidates, which as commented in the introduction include the $K_0^*(800)$, are relatively close. In particular, the $K_0^*(800)$, 
having  a width of the order of 600 MeV and a pole mass around 700 MeV,
is less than one width and a half away from the $K^*_0(1430)$.

Moreover, since the $K^*_2(1980)$ 
seems to be a good candidate for the next partner of the $K_0^*(1430)$, 
we can estimate $\Gamma_{K^*_2(1980)\rightarrow K\pi}\simeq 97\,$MeV
by approximating the width as $\Gamma={\text Im}\alpha/(M \re \alpha')$, 
assuming $K^*_2(1980)$ can be described by a Breit-Wigner form. 
Unfortunately, no estimate of this
partial width is given in the RPP, but at least
our result is smaller
than the total width $\Gamma_{tot}=(373\pm33\pm60)\,$MeV.

As a final remark on Fig.\ref{Fig:alphak0}, we want to comment on the apparent 
cusp seen at threshold 
in the $K_0^*(1430)$ trajectory, even if it lies beyond the strict applicability region
of our method (since it lies in the area covered with a mesh).
It is just an artifact of our approximation due to our assumption that the Regge pole dominates the amplitude. 
But as seen in Fig.\ref{Fig:ampliK0}, for the $K_0^*(1430)$ this dominance is only a good approximation
 in an energy region of the order of the resonance width around the nominal mass.
Being a Regge pole in the complex-$l$ plane $\beta(s)$ must carry a $q^{2\alpha(s)}$ factor.  Right at the pole 
this becomes exactly $q^{2l}$, as expected from partial wave kinematics. However, the pole does not dominate at threshold,
where $\alpha(s_{th})\neq l$. Therefore at threshold the approximation $q^{2l}\sim q^{2\alpha}$ is not so good.

Now, for trajectories of scalar particles Re$\,\alpha$ is negative between threshold and the pole.
Consequently, in the first step of the calculation, $q^{2\alpha}$ diverges at threshold and so does $\beta(s)$.
If $0>$Re$\,\alpha>-0.5$ this spurious divergence is compensated in Im$\alpha=\rho(s) \beta(s)$ by the $\rho(s)\sim q$
factor. Thus the artifact due to extending
our approximation to threshold goes unnoticed. 
This will be the case of the light $K_0^*(800)$ (or the $\sigma$ resonance studied in \cite{Londergan:2013dza}).
However, for the $K_0^*(1430)$, Re$\,\alpha$ can become close or smaller
than $-0.5$ at threshold, making Im$\,\alpha\rightarrow +\infty$ there. 
But recall that our equations are solved iteratively.
Then, if at any step of the calculation we feed in the equations a huge positive 
Im$\,\alpha(s')$ near threshold, the resulting Re$\,\alpha(s)$ changes sign
becoming positive near threshold. With further iterations
the solution at threshold always stabilizes at values of Re$\,\alpha(s)>-0.5$. Beyond threshold it can be negative 
until it reaches the value of $\alpha=0$ at the resonance mass.
Therefore the spurious behavior of the trajectory 
around $s=s_{th}$ is not due to the presence of another Regge pole like the $K_0^*(800)$,
 but just to assuming that at threshold the amplitude is dominated by the $K_0^*(1430)$ pole.

One might then worry that this artifact may spoil our calculation, but we have also checked explicitly that the part of the integral 
around threshold is negligible
for the result of the trajectory in the applicability region. As explained above we could have restricted the integrals
to the region around the resonance and the result would have changed little.

\subsection{$K^*_0(800)$ resonance}
\label{subsec:K0(800)}
This is a very interesting state, because, as commented in the introduction, it is a firm candidate to 
be a non-ordinary meson together with the other members of the light scalar nonet.
There is also a longstanding debate on its parameters and even its very existence, and in the RPP it is still listed as "Needs Confirmation". However, all sensible implementations of chiral symmetry and unitarity obtain a pole for this state, which is also necessary for the understanding of several heavy meson decays (see, for instance \cite{Ablikim:2010ab}). Within unitarized chiral perturbation theory 
it was shown  that this state does not follow the $N_c$ behavior of ordinary mesons \cite{Pelaez:2003dy,Pelaez:2004xp} and that for heavy quark masses it would become a virtual state \cite{Nebreda:2010wv}, which has been recently confirmed on the lattice \cite{Dudek:2014qha}. The most rigorous determination of its parameters was obtained from the dispersive analysis in \cite{DescotesGenon:2006uk} using the Roy-Steiner equations with unitarity and low-energy chiral constraints. 
In that work the pole position is given explicitly, but unfortunately not the residue, which is needed for our approach. For this reason we will use the parameters  obtained in \cite{Pelaez:2016tgi}, in which a conformal expansion with the correct analytic properties was fitted to $K\pi$ scattering data constrained to satisfy forward dispersion relations up to 1.6 GeV. The pole parameters we will use are thus
\begin{eqnarray}
\sqrt{s_{\kappa}}&=&(680\pm15)-i(334\pm8)\text{ MeV},\nonumber\\
 |g_{\kappa}|^2&=&25.0\pm0.6\text{ GeV}^{2},
\label{kappacfd}
\end{eqnarray} 
which are fairly consistent with the position provided in \cite{DescotesGenon:2006uk} and the RPP. 

\begin{figure*}
\hspace*{-.6cm}
\centering
\centerline{\includegraphics[width=0.48\linewidth]{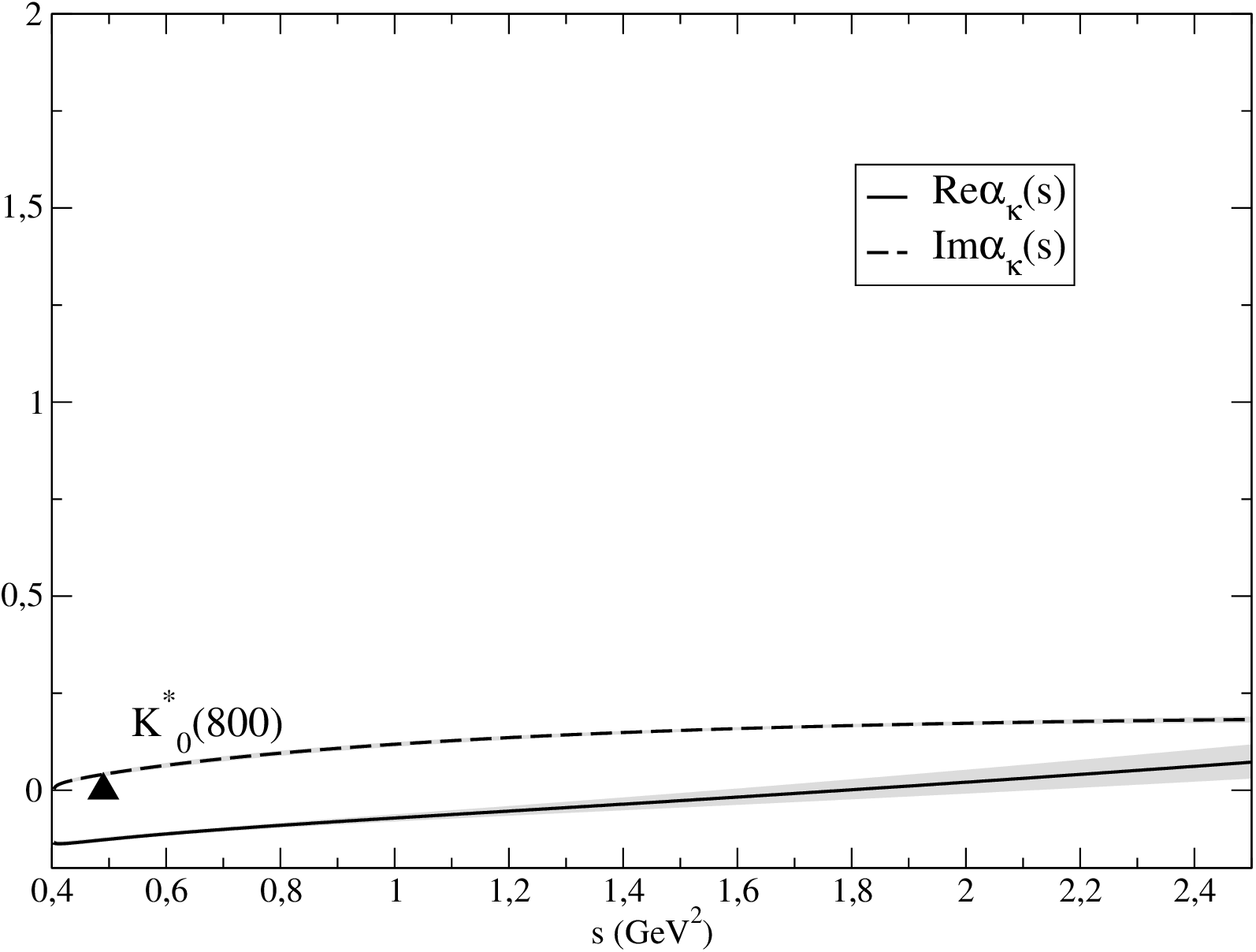}   \includegraphics[width=0.48\linewidth]{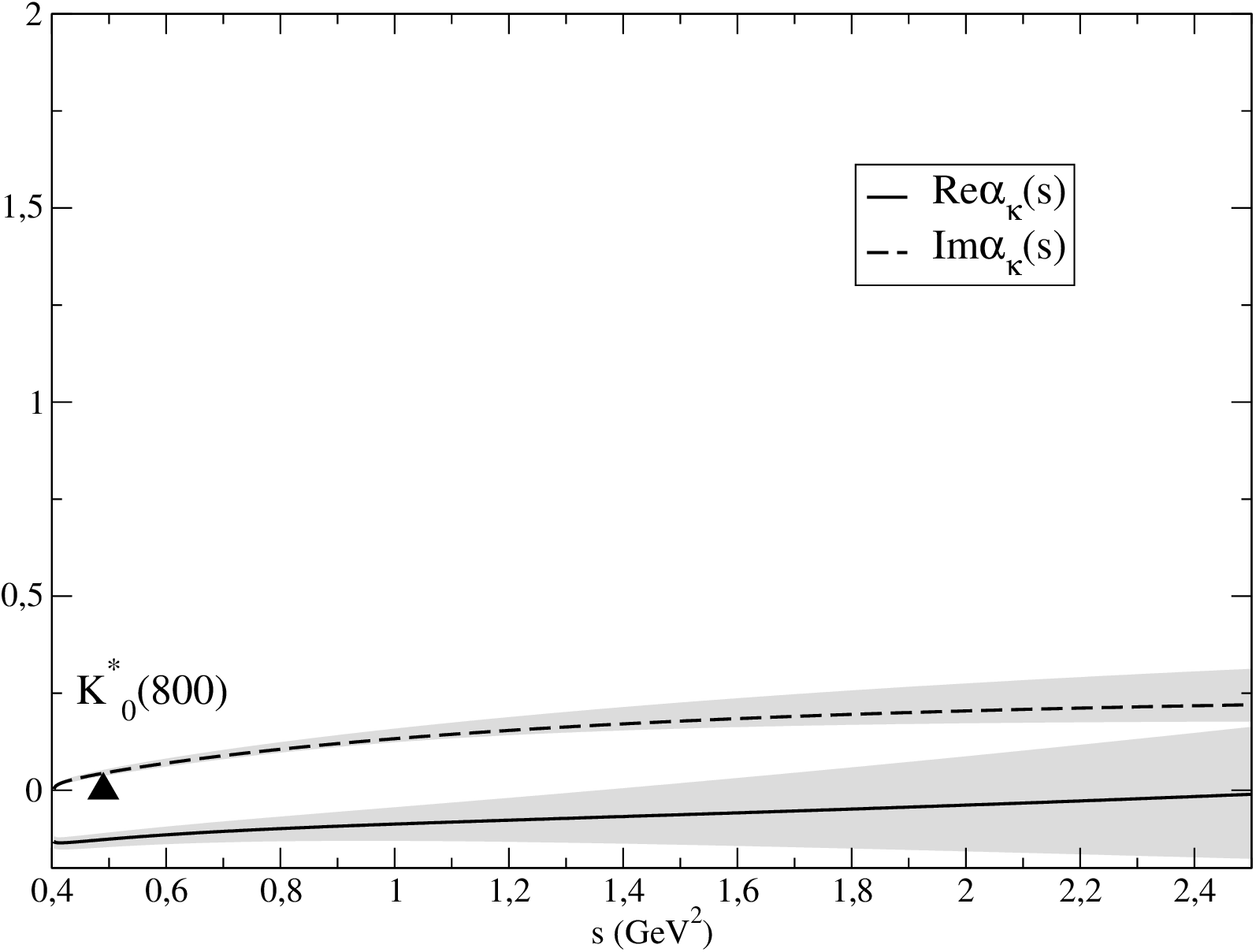}}
 \caption{\rm \label{Fig:alphakappa} 
Real (solid) and imaginary (dashed) parts of the $K^*_0(800)$ Regge trajectory for both FDR \cite{Pelaez:2016tgi} and PFDR \cite{Pelaez:2016klv} results.
 The real part is smaller than the imaginary part in the whole energy region. The slope of the Regge trajectory is almost one order of magnitude smaller than the usual ones. In addition there cannot be any candidate for this resonance since the real part is below 0.25 up to $s=5$ GeV$^{-2}$. }
\end{figure*}

As in previous sections, the pole parameters above are then fitted with our Regge amplitude
in Eq.\eqref{reggepw}, neglecting the background and 
with the Regge slope and residue satisfying the dispersive representation in Eqs.  \eqref{iteration1} to \eqref{betafromalpha}. The pole obtained from this fit is located at $\sqrt{s_{\kappa}}=(680\pm15)-i(334\pm8)\text{ MeV}$, with a coupling $|g_{\kappa}|^2=25.1\pm0.5\text{ GeV}^{2}$, very consistent with the input values. The parameters of the fit are
\begin{align}
&\alpha_0=0.28\pm0.02\,;\hspace{0.3 cm}
\alpha'=0.15\pm0.03 \text{ GeV}^{-2};\\ \nonumber
&b_0=0.44\pm0.04\text{ GeV}^{-2}\,,\label{eq:paramskappa}
\end{align}
and the corresponding trajectory $\alpha(s)$ is shown in the left panel
of Fig.~\ref{Fig:alphakappa}. It is clearly not an ordinary Regge trajectory, since it is not predominantly real, the real part is non linear and the slope (at the $K^*_0(800)$ mass) is almost one order of magnitude smaller than the usual $\alpha'\simeq 0.9$ GeV$^{-2}$ slope for ordinary mesons.

\begin{figure*}
\centerline{\includegraphics[width=0.47\linewidth]{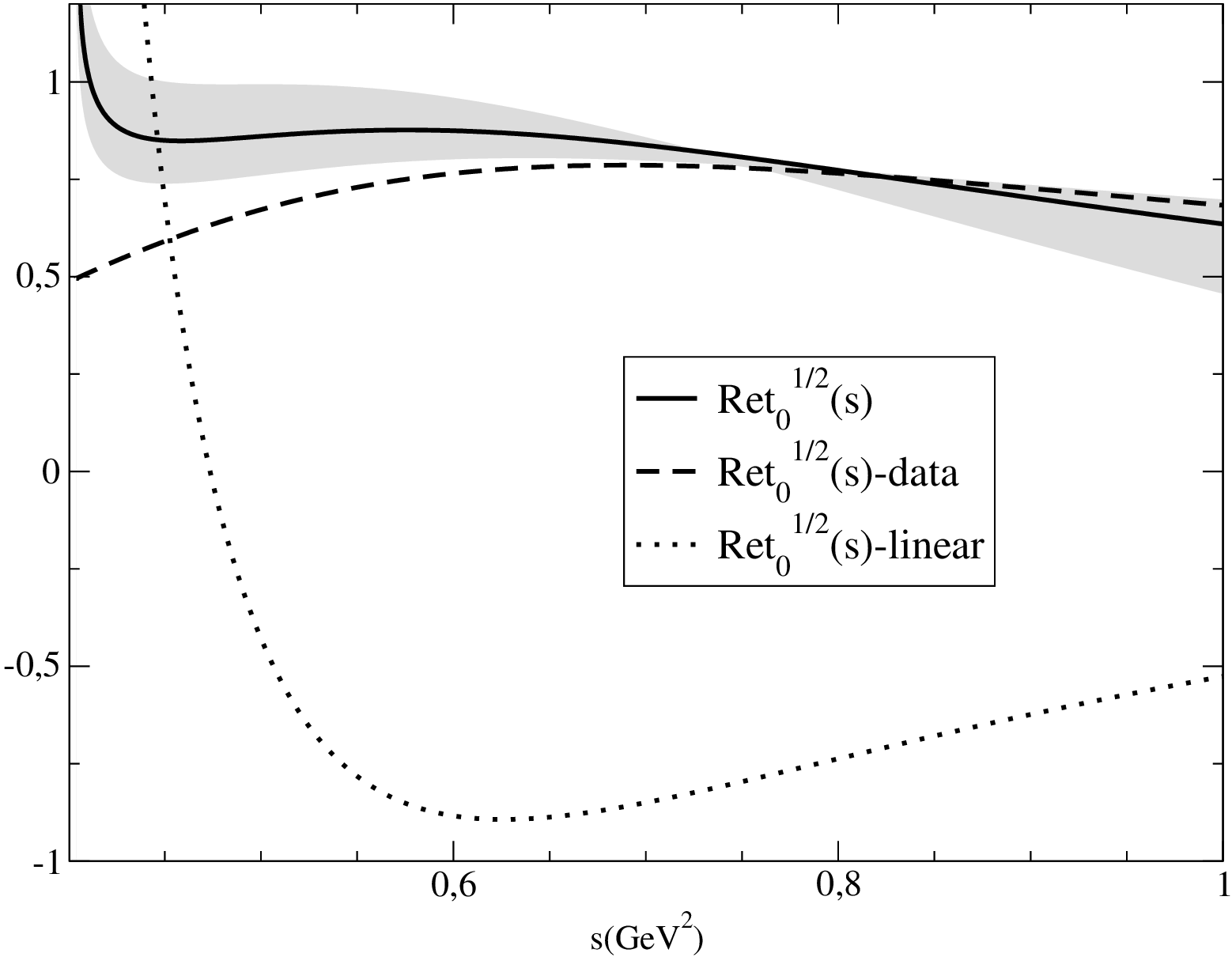} \hspace*{.6cm}  \includegraphics[width=0.47\linewidth]{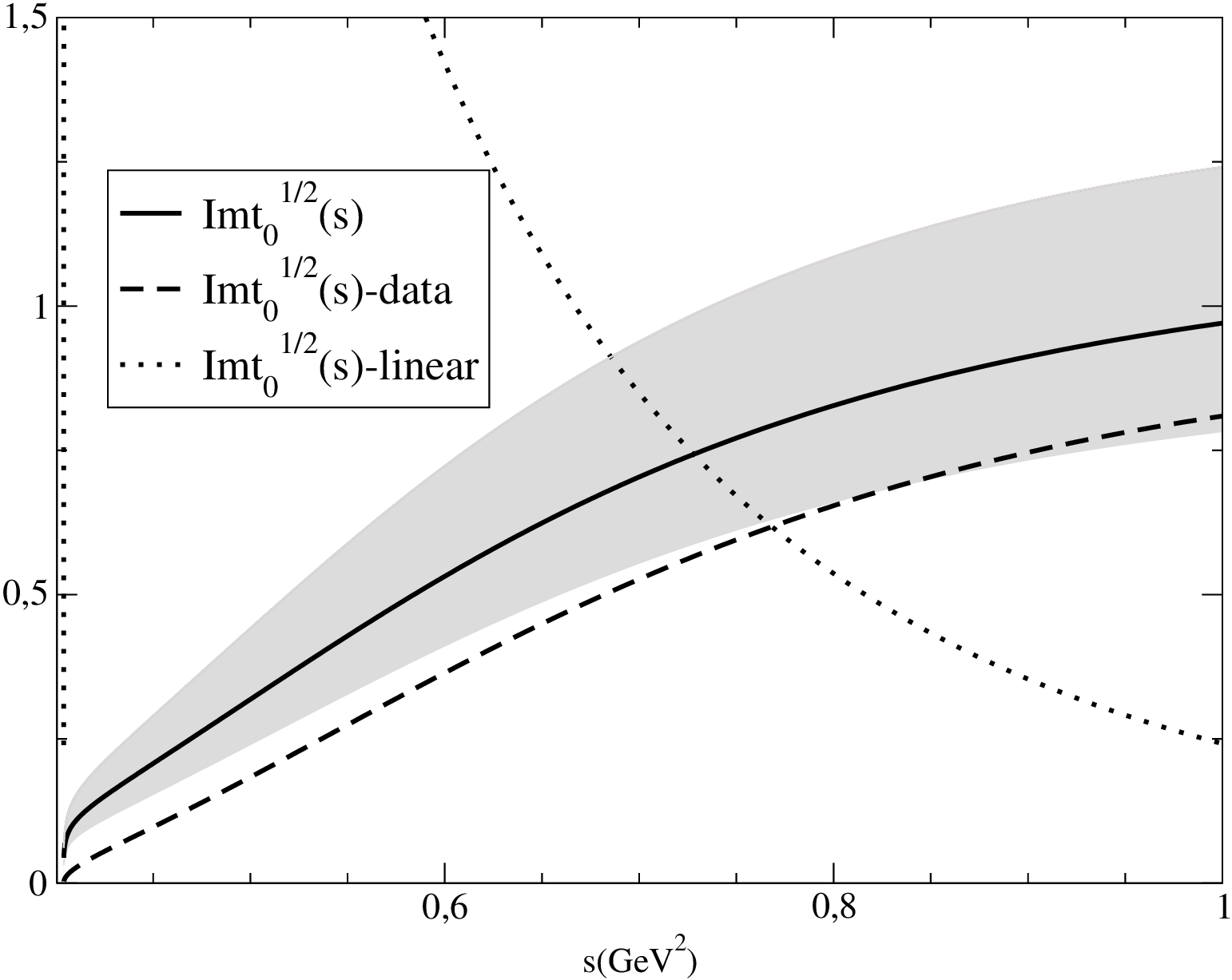}}
 \caption{\rm \label{Fig:reggekappa} 
Real and imaginary parts of the partial wave $t^{1/2}_0(s)$. Dashed lines represent the constrained fit to data in \cite{Pelaez:2016tgi}. The solid curves represent the single Regge-pole partial wave determined using the dispersive representation of the Regge trajectory. The estimated uncertainties are shown as gray bands. The dotted lines, which are completely at odds with the data curve, represent the single Regge-pole partial wave when imposing 
an ordinary linear Regge trajectory with $\alpha'\simeq 0.9$ GeV$^{-2}$, as a solution from the dispersive equations of $\alpha(s)$ and $\beta(s)$. }
\end{figure*}

In Fig.\ref{Fig:reggekappa} 
we compare the one-pole partial wave of Eq. \eqref{reggepw},  
when the pole follows this non-ordinary trajectory (continuous line and dark error band), with the
dispersive data fit from \cite{Pelaez:2016tgi} (dashed line).
It can be seen that the non-ordinary Regge-pole amplitude consistently
dominates the amplitude in that region, even though we have only fitted the pole position and residue deep in the complex plane.

To check the robustness of our results we have performed some further tests. First,
we have also used the pole obtained in a recent work \cite{Pelaez:2016klv}, which also used the constrained data analysis of \cite{Pelaez:2016tgi}, although the pole was not obtained from the conformal fit there, but from a sequence of Pad\'e approximants. The 
pole parameters thus obtained were 
\begin{eqnarray}
\sqrt{s_{\kappa}}&=&(670\pm18)-i(295\pm28)\text{ MeV}, \nonumber\\
 |g_{\kappa}|^2&=&20.0^{+3.7}_{-3.4}\text{ GeV}^{2}.
\label{eq:kappapar}
\end{eqnarray}
 Note that the uncertainties in these parameters are more conservative than those in Eq. \eqref{kappacfd}. Once again we fit this pole with our single-Regge-pole amplitude. The resulting pole is at  $\sqrt{s_{\kappa}}=(670\pm18)-i(295\pm28)\text{ MeV}$, with $|g_{\kappa}|^2=20.0^{+3.7}_{-2.2}\text{ GeV}^{2}$, 
almost identical-2 to the input. The parameters obtained for the Regge trajectory associated to this pole are
\begin{align}
&\alpha_0=0.27\pm0.03\,;\hspace{3mm} 
\alpha'=0.11\pm 0.9 \text{ GeV}^{-2}; \\ \nonumber
&b_0=0.45^{+0.11}_{-0.8}\text{ GeV}^{-2}\,,\label{eq:paramskappacfd}
\end{align} 
very consistent with the 
determination in Eq. \eqref{eq:paramskappa}, although more conservative. 
Once again 
we see on the right panel of Fig.\ref{Fig:alphakappa} that 
the resulting trajectory is very different from that expected for an ordinary meson and 
very consistent with the trajectory in the left panel, although with more conservative error bands.

As a second test, we have performed the same analysis but imposing an ordinary slope
$\alpha'=0.9\,$GeV$^{-2}$. Despite having one less free parameter for the fit, 
it is still possible to fit the pole position fairly well, finding
$\sqrt{s_{\kappa}}=683-i331\text{ MeV}$, with the coupling $|g_{\kappa}|^2=25.1\text{ GeV}^{2}$. 
With $\alpha'$ fixed we now find a linear Regge trajectory and one could 
be tempted to think that we could also consider the $\kappa$ to lie in an ordinary Regge trajectory.
However, the resulting amplitude may describe the pole, but fails completely to describe
in the real axis
the amplitude fitted to data. This can be seen in Fig.\ref{Fig:reggekappa},
where we show as dotted lines the real and imaginary parts of the resulting amplitude when imposing an ordinary linear Regge trajectory for the $K_0^*(800)$, versus the partial wave fitted to data. 
Therefore the linear Regge trajectory with universal slope does not yield a pole 
that dominates the observed amplitude.

Moreover, if one was to assume an ordinary linear Regge trajectory $J=\alpha_0+\alpha' M^2$
for the $K_0^*(800)$ with the universal value $0.9\,$GeV$^{-2}$, then taking $M_\kappa\simeq 0.68\,$GeV,
one finds $\alpha_0\simeq-0.42$. 
Hence the first partner with $J=2$ in this trajectory would appear at $1.64\,$GeV.
No $J^P=2^+$ resonance is identified in the RPP with such a mass. 
The closest one is the $K_2^*(1430)$, but that would require $\alpha'=1.26\,$ GeV$^{-2}$, 
very inconsistent with the universal slope. The second closest $J^P=2^+$ resonance
is the $K_2^*(1980)$, but we have already seen in the previous section that this one would fit
better in the $K_0^*(1430)$ trajectory. Actually, to make it the partner of the $K_0^*(800)$ 
in a linear trajectory, a value of $\alpha'\simeq 0.58\,$GeV$^{-2}$ is required, also rather different from
the universal value. A $K(1630)$ resonance is listed in the RPP with unknown $J^P$, but it is 
not confirmed by more than one experiment, it is omitted from the summary tables and has a surprisingly small width of $16^{+19}_{-16}\,$MeV, which makes its existence very questionable.
Furthermore, even in QCD-inspired quark models \cite{Godfrey:1985xj}, only two $J^P=2^+$
states are listed below 2 GeV and they can be nicely identified with the $K_2^*(1430)$ and $K_2^*(1980)$.
Therefore, even from more familiar 
phenomenology there is no natural candidate for a partner of the $K_0^*(800)$ 
if it lies in an ordinary trajectory.

A third test is that we have also tried to 
fit the pole without factorizing the Adler zero, 
but once again the result is at odds with the data. 

\begin{figure*}
\begin{center}
\includegraphics[scale=0.4]{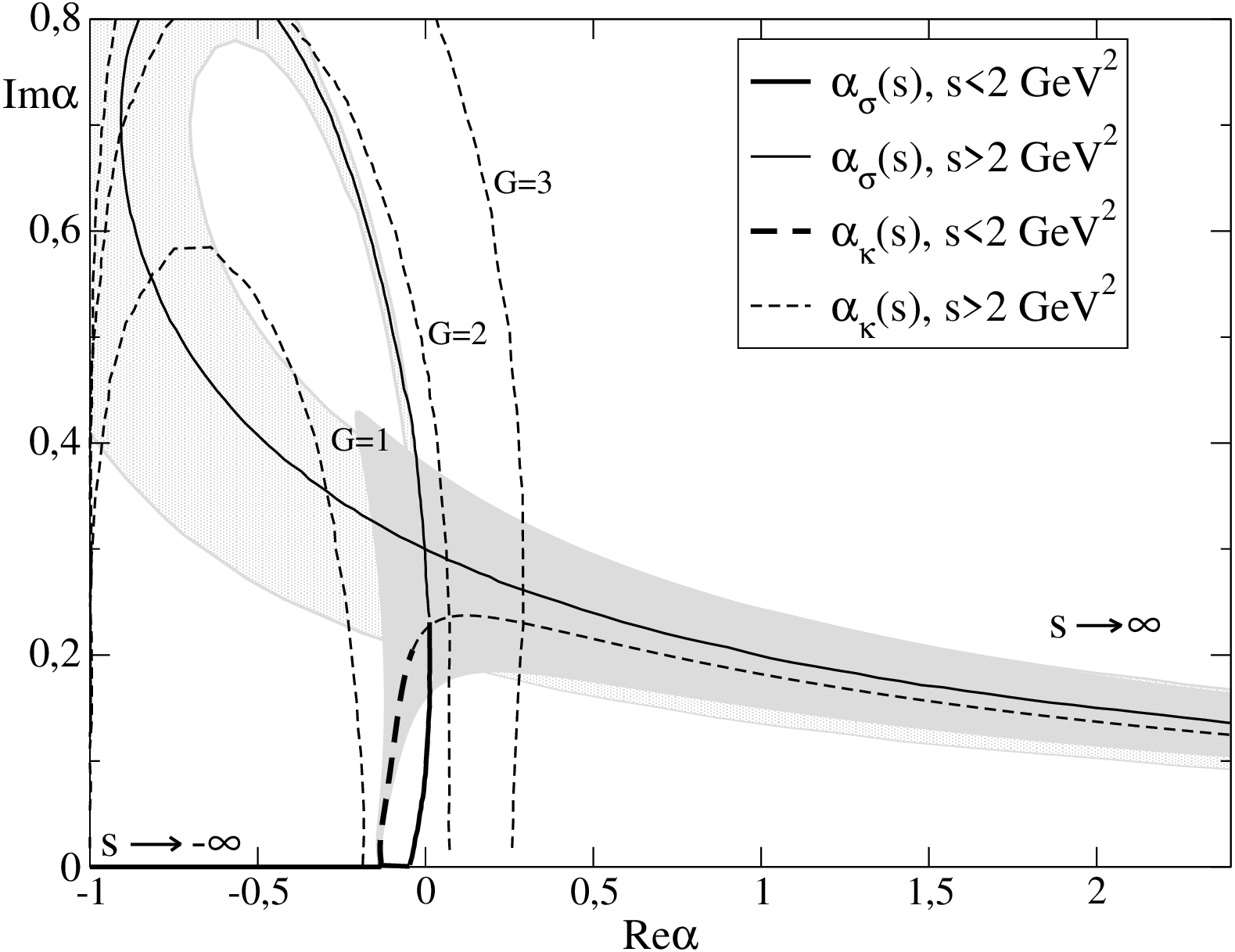}
 \caption{\rm \label{Fig:yukawacomp} 
At low and intermediate energies (thick lines), the trajectories of the $\sigma$ and the $\kappa$ are similar
to those of Yukawa potentials \cite{yukawamodels} $V(r)= Ga \exp(-r/a)/r$ (thin dashed lines labeled with 
different values of $G$).
Beyond 2 GeV$^2$, we plot our results as thin lines because they should
be considered just as extrapolations.}
\end{center}
\end{figure*}

These results strongly support the non-ordinary nature of the $K^*_0(800)$ resonance, or $\kappa$ meson.
Furthermore, a rather similar non-ordinary Regge behavior has been recently observed 
\cite{Londergan:2013dza}
for the $f_0(500)$ resonance, formerly known as the $\sigma$ meson. 
As commented in the introduction, there is a  rather general agreement in the literature that these two resonances would belong to the same light scalar multiplet.
The similarities between the trajectories of these two states is shown in Fig.~\ref{Fig:yukawacomp} where we plot $\im \alpha(s)$ versus $\re \alpha(s)$ for the $K^*_0(800)$ together with the results obtained for the $\sigma$ in \cite{Londergan:2013dza}. It should be noted that for the 
$f_0(500)$ 
a very robust behavior below 2 GeV$^2$ was found, which is qualitatively similar to the one we also find for the $K_0^*(800)$. 
Namely, both the $f_0(500)$  and $K^*_0(800)$ trajectories in this plane are almost exactly real up to 
a value of $\re \alpha(s)$ between $-0.5$ and 0, where 
both curves rise almost vertically developing an imaginary part 
up to slightly above 0.2, without barely changing the real part. 
This happens for values of $s<2\,$GeV$^2$, where we expect our method to be valid,
 and we have plotted this part of the curves with thick lines.
It is worth noting that this is the typical behavior of Regge trajectories of Yukawa potentials $V(r)= Ga \exp(-r/a)/r$ at low energies \cite{yukawamodels}, which we have plotted as dashed lines for different values of $G$. Therefore, it seems that both the $\sigma$ and the $\kappa$ mesons have a Regge trajectory at low energies that is qualitatively similar to Yukawa potential trajectories.

Above 2 GeV$^{2}$ our method becomes less reliable, but we still show the results as thin  curves
for completeness. It should be considered just a mere extrapolation. 
Moreover, given the high energies under consideration, the comparison with non-relativistic Yukawa potential does not make
sense any longer. The sigma presents two possible behaviors, in one of them $\im \alpha$ reaches a value between 0.2 and 0.4 and then decreases slowly while the real part starts increasing again. This is the same behavior we find for the $K_0^*(800)$. However, within uncertainties,
the $f_0(500)$ has another possible behavior which still follows the Yukawa trajectory above 2 GeV$^2$. In any case, at those high energies these trajectories are still non-ordinary as seen in Fig.~\ref{Fig:alphakappa} for the $K_0^*(800)$.

Once the semiquantitative analogy with Yukawa potentials has been established, 
it is possible to estimate the Yukawa parameters that mimic best the $\sigma$ and $\kappa$ trajectories. The trajectory of the $f_0(500)$ is almost equal to a $G=2$ curve up to $s=2$ GeV$^2$, 
while the curve with $G=1.4$ is rather similar to the $K_0^*(800)$ trajectory. Using the parameterizations 
of Yukawa Regge trajectories in \cite{yukawamodels} we can estimate the effective ranges 
of the Yukawa potential in the $\sigma$ case \cite{Londergan:2013dza}: $a_{\pi \pi}=0.5$ GeV$^{-1}\simeq 0.1\,$fm, as well as in the $\kappa$ case: $a_{\pi K}\simeq0.36$ GeV$^{-1}\simeq 0.07\,$fm.

The range of the interaction is a relevant quantity because there is some interest in the literature in determining the size of resonances and whether they are compact or extended objects. If they are extended the "molecular" interpretation, i.e. predominantly formed from a 
two-meson interaction, would be preferred over the interpretation where the binding force is between quarks.
Unfortunately, the concept of size is poorly defined for resonances, since the spatial part of their wave-function is non-normalizable. In particular, the simple extrapolation of methods that can determine the compositeness of bound states \cite{weinbergdeuterium} is not directly translatable to resonances. Some efforts to generalize the concept of size, radius or some compositeness criteria  can be found in \cite{compositenesscriteria,Albaladejo:2012te}. For instance, in \cite{Albaladejo:2012te}, the scalar radius was generalized to a complex number finding: $\langle r^2\rangle_s^\sigma=(0.19\pm0.02)-i(0.06\pm002)\,$fm$^2$, i.e., close in modulus but smaller than a typical meson radius.
In contrast, in our case the range of the Yukawa potential that mimics the Regge trajectory is a well-defined quantity, giving a very intuitive 
picture of the range of the interactions responsible for the formation of the resonance.

Returning to our values, the interaction range we have found is somewhat smaller but
of the order of typical values of meson-meson
observables like the scalar scattering lengths $a_0^{(I)}$, where $I$ is the isospin in $\pi\pi$ or $\pi K$. Their values are $a_0^{(0)}\simeq a_0^{(1/2)}\simeq1.6$ GeV$^{-1}\simeq 0.3\,$fm. 
Therefore the range of interactions producing the $\kappa$ and $\sigma$ 
seem comparable but somewhat smaller than meson-meson interactions themselves. 
Keeping in mind that the range of the interaction is not directly the ``size'' of a resonance
our interaction ranges compare rather well with the modulus of the radius
$\sqrt{\vert \langle r^2\rangle_s^\sigma\vert}\simeq 0.45\,$fm
obtained in \cite{Albaladejo:2012te}. Very naively one would  expect 
the interaction range to be smaller than any generalization of the radius, since after all
the resonance is a quasi-bound state that escapes from the typical interaction range.

Moreover, if the interactions is of the meson-meson type, 
a rather natural  mass scale for the system is the reduced mass of the two mesons. 
If a naive $a\sim 1/\mu$ proportionality is assumed for the range, one would expect: 
$a_{\pi \pi}/a_{\pi K} \approx \mu_{\pi K}/\mu_{\pi \pi}=1.56$.
Remarkably, from our previous estimates of the effective range we find 
 $a_{\pi \pi}/a_{\pi K}\simeq 1.39$, i.e. within a 10\% from that expectation. 
This would also be consistent with the interpretation that 
 both the $\sigma$ and the $\kappa$ are predominantly meson-meson resonances.

\section{Conclusions}
\label{conclusions}

In contrast to the usual phenomenological approach of fitting 
the spin and squared mass of hadrons into linear trajectories, 
in this work we have applied a method {\it to calculate} Regge trajectories
without assuming a priori their functional form.
In addition, instead of using as input the parameters of several resonances,
the only input is the position and residue of the pole associated to a single resonance.
The method applies to elastic meson resonances, i.e., 
those resonances that decay almost completely into a single two-meson channel.
In particular, the method has been previously shown to predict 
that the $\rho(770), f_2(1270), f_2'(1525)$ and $K^*(892)$ Regge trajectories
are  almost real and linear with a constant slope of roughly 0.9 GeV$^{-2}$,
in good agreement with the expectations for a confining interaction
between a constituent quark and an antiquark, i.e. for ordinary mesons.
In contrast, the Regge trajectory of the controversial $f_0(500)$
or $\sigma$ meson, was found to be non-real, not linear and with a much smaller slope 
than ordinary trajectories.

Here we have applied this method to the controversial $K_0^*(800)$ or $\kappa$ meson
and to the almost elastic and scalar strange $K_0^*(1430)$ resonance. 
For the latter we have found a rather ordinary trajectory which suggests that its
nature is largely dominated by confining quark-antiquark interactions. The $K^*_2(1980)$ is even a 
fairly reasonable candidate to be its next trajectory partner with $J=2$, although 
with some tension in the parameters, so that it would not be too surprising if
the $K_0^*(1430)$ had other subdominant, but sizable, non-$q\bar q$ components.

Of course,  the most interesting result of this work is the trajectory of the controversial $K_0^*(800)$, which for long has been considered a non-ordinary meson candidate.
The Regge trajectory we find for this resonance is not predominantly real and 
its real part is not linear. This clearly supports the identification of this state as a non-ordinary meson. Moreover, its Regge trajectory slope at the physical mass is much smaller than the 
universal slope of ordinary trajectories. This also seems to suggest that meson physics, more than interquark interactions, might be responsible for its formation.

In addition, the trajectory of the $K_0^*(800)$ is very similar to that already found for the $f_0(500)$ or $\sigma$ meson, 
thus supporting the widely extended view that both belong to the same light scalar nonet.
Furthermore, at low energies the trajectories of these two resonances have very significant similarities
with the trajectories of Yukawa potentials between two mesons, whose range has been estimated here:  
$\sim$ 0.36 GeV$^{-1}$ for the $\kappa$ and $\sim$ 0.5 GeV$^{-1}$ for the $\sigma$. 
This is of interest because, being non-normalizable states,
 it is very hard to define the concept of ``size'' for resonances, whereas the range of the 
interaction that produces the resonance is a well-defined and intuitive concept. 
Incidentally, the interaction range seems compatible with 
 a scaling inversely proportional to the reduced mass of the meson system, 
which also seems relatively natural if the meson-meson interactions plays a dominant role in the 
resonance formation.

Altogether, our results seem to support a predominantly non-ordinary nature for the $K_0^*(800)$ and suggest that its formation is mainly due to meson-meson dynamics.

\section*{Acknowledgments} 

Work supported by the Spanish Projects  FPA2014-53375-C2-2,
FPA2016-75654-C2-2-P and the group UPARCOS
and the Spanish Excellence network HADRONet FIS2014-57026-REDT.
A. Rodas would also like to acknowledge the financial support of 
the Universidad Complutense de Madrid through a predoctoral scholarship.


\vspace*{-.2cm}
\begin{thebibliography}{99}
\vspace*{-.2cm}




\bibitem{Londergan:2013dza} 
  J.~T.~Londergan, J.~Nebreda, J.~R.~Pelaez and A.~Szczepaniak,
  Phys.\ Lett.\ B {\bf 729}, 9 (2014).
	
\bibitem{Carrasco:2015fva} 
  J.~A.~Carrasco, J.~Nebreda, J.~R.~Pelaez and A.~P.~Szczepaniak,
  Phys.\ Lett.\ B {\bf 749}, 399 (2015).



\bibitem{Reggeintro} 
P.B.D Collins, {\em An Introduction to Regge Theory and High Energy Physics.}
Cambride Univerity Press, Cambridge (1977). V. M. Gribov, {\em The Theory of Complex Angular Momenta.} Cambride Univerity Press, Cambridge (2003).

\bibitem{Collins-PLB} P.D.B.~Collins, R.C.~Johnson, E.J.~Squires, Phys.\ Lett. B{\bf 26}, 223 (1968). 

\bibitem{Chu:1969ga} 
G. Epstein and P. Kaus, Phys. Rev. {\bf 166}, 1633 (1968);
  S.~-Y.~Chu,
  G.~Epstein, P.~Kaus, R.~C.~Slansky and F.~Zachariasen,
  Phys.\ Rev.\  {\bf 175}, 2098 (1968).



\bibitem{Caprini:2005zr} 
  I.~Caprini, G.~Colangelo and H.~Leutwyler,
  Phys.\ Rev.\ Lett.\  {\bf 96}, 132001 (2006).
  [hep-ph/0512364].
  R.~Garcia-Martin, R.~Kaminski, J.~R.~Pelaez and J.~Ruiz de Elvira,
  Phys.\ Rev.\ Lett.\  {\bf 107}, 072001 (2011).

\bibitem{Anisovich:2000kxa} 
  A.~V.~Anisovich, V.~V.~Anisovich and A.~V.~Sarantsev,
  Phys.\ Rev.\ D {\bf 62}, 051502 (2000).


\bibitem{Pelaez:2015qba} 
  J.~R.~Pelaez,
  Phys.\ Rept.\  {\bf 658}, 1 (2016).

\bibitem{RPP}
C. Patrignani et al. (Particle Data Group), Chin. Phys. C, 40, 100001 (2016). 

\bibitem{Jaffe:1976ig} 
  R.~L.~Jaffe,
  Phys.\ Rev.\ D {\bf 15}, 267 (1977).

\bibitem{vanBeveren:1986ea} 
  E.~van Beveren,{\it et al.} 
  Z.\ Phys.\ C {\bf 30}, 615 (1986).

\bibitem{Oller:1997ng}
J.~A.~Oller, E.~Oset and J.~R.~Pelaez,
Phys.\ Rev.\ Lett.\  {\bf 80} (1998) 3452;
  Phys.\ Rev.\ D {\bf 59}, 074001 (1999)
  [Erratum-ibid.\ D {\bf 60}, 099906 (1999)]
  [Erratum-ibid.\ D {\bf 75}, 099903 (2007)]

\bibitem{Black:1998zc} 
  D.~Black, A.~H.~Fariborz, F.~Sannino and J.~Schechter,
  Phys.\ Rev.\ D {\bf 58}, 054012 (1998);
  Phys.\ Rev.\ D {\bf 59}, 074026 (1999).
%

\bibitem{Oller:1998zr}
  J.~A.~Oller and E.~Oset,
  Phys.\ Rev.\ D {\bf 60} (1999) 074023.

\bibitem{Close:2002zu} 
  F.~E.~Close and N.~A.~Tornqvist,
  J.\ Phys.\ G {\bf 28}, R249 (2002).

\bibitem{Pelaez:2003dy}
  J.~R.~Pelaez,
  Phys.\ Rev.\ Lett.\  {\bf 92}, 102001 (2004).

\bibitem{Pelaez:2004xp} 
  J.~R.~Pelaez,
  Mod.\ Phys.\ Lett.\ A {\bf 19}, 2879 (2004).


	
\bibitem{DescotesGenon:2006uk}
  S.~Descotes-Genon and B.~Moussallam,
  Eur.\ Phys.\ J.\ C {\bf 48} (2006) 553.

\bibitem{Wolkanowski:2015jtc} 
  T.~Wolkanowski, M.~Soltysiak and F.~Giacosa,
  Nucl.\ Phys.\ B {\bf 909}, 418 (2016).
	
	

\bibitem{Cherry:2000ut} 
  S.~N.~Cherry and M.~R.~Pennington,
  Nucl.\ Phys.\ A {\bf 688}, 823 (2001).


\bibitem{Pelaez:2016klv} 
  J.~R.~Pelaez, A.~Rodas and J.~Ruiz de Elvira,
  Eur.\ Phys.\ J.\ C {\bf 77}, no. 2, 91 (2017).


\bibitem{Dudek:2014qha} 
  J.~J.~Dudek {\it et al.}  [Hadron Spectrum Collaboration],
  Phys.\ Rev.\ Lett.\  {\bf 113}, no. 18, 182001 (2014).


  
\bibitem{Nebreda:2010wv}
  J.~Nebreda and J.~R.~Pelaez.,
  Phys.\ Rev.\  D {\bf 81}, 054035 (2010).
  


\bibitem{Pelaez:2016tgi} 
  J.~R.~Pelaez and A.~Rodas,
  Phys.\ Rev.\ D {\bf 93}, no. 7, 074025 (2016).




\bibitem{Kpiscattering}
  P.~Estabrooks {\it et al.}, 
  Nucl.\ Phys.\ B {\bf 133}, 490 (1978).
  D.~Aston {\it et al.},
  Nucl.\ Phys.\ B {\bf 296}, 493 (1988).

 
\bibitem{Fariborz:2005gm} 
  A.~H.~Fariborz, R.~Jora and J.~Schechter,
  Phys.\ Rev.\ D {\bf 72}, 034001 (2005).
  Phys.\ Rev.\ D {\bf 76}, 014011 (2007).
  Phys.\ Rev.\ D {\bf 77}, 034006 (2008).
  A.~H.~Fariborz, R.~Jora, J.~Schechter and M.~N.~Shahid,
  Phys.\ Rev.\ D {\bf 84}, 113004 (2011).


\bibitem{Giacosa:2006tf} 
  F.~Giacosa,
  Phys.\ Rev.\ D {\bf 75}, 054007 (2007).


\bibitem{Ablikim:2010ab} 
  M.~Ablikim {\it et al.} [BES Collaboration],
  Phys.\ Lett.\ B {\bf 698}, 183 (2011).


\bibitem{Godfrey:1985xj} 
  S.~Godfrey and N.~Isgur,
  Phys.\ Rev.\ D {\bf 32}, 189 (1985).


 
\bibitem{yukawamodels} 
  C.~Lovelace and D.~Masson,
	Nuovo Cimento {\bf 26} 472 (1962).
	%
	A.O.~Barut and F.~Calogero,
	Phys.\ Rev. {\bf 128} 1383 (1962).
	%
	A.~Ahamadzadeh, P.G.~Burke and C.~Tate,
	Phys.\ Rev. {\bf 131} 1315 (1963).


\bibitem{weinbergdeuterium}
S. Weinberg, Phys. Rev. {\bf
130}, 776 (1963); {\bf 131}, 440 (1963); B{\bf 137}, 672 (1965).

\bibitem{compositenesscriteria}
  V.~Baru, J.~Haidenbauer, C.~Hanhart, Y.~Kalashnikova and A.~E.~Kudryavtsev,
  Phys.\ Lett.\ B {\bf 586}, 53 (2004).
  J.~Yamagata-Sekihara, J.~Nieves and E.~Oset,
  Phys.\ Rev.\ D {\bf 83}, 014003 (2011).
  T.~Hyodo, D.~Jido and A.~Hosaka,
  Phys.\ Rev.\ C {\bf 85}, 015201 (2012).
  F.~Aceti, L.~R.~Dai, L.~S.~Geng, E.~Oset and Y.~Zhang,
  Eur.\ Phys.\ J.\ A {\bf 50}, 57 (2014).
  Z.~H.~Guo, J.~A.~Oller,
  Phys.\ Rev.\ D {\bf 93}(9),  096001 (2016).



\bibitem{Albaladejo:2012te} 
  M.~Albaladejo, J.~A.~Oller,
  Phys.\ Rev.\ D {\bf 86}, 034003 (2012).







\end{thebibliography}
\end{document}